\documentclass[superscriptaddress,twocolumn,showpacs,pre]{revtex4}
\newcommand{\figurewidth}{\columnwidth}
\usepackage{graphicx,amsmath}

\begin{document}

\title{
Complexity of several constraint satisfaction problems using the heuristic,
classical, algorithm, WalkSAT}

\author{Marco ~Guidetti}
\affiliation{Dipartimento di Fisica, Universit\`a di Ferrara and INFN---Sezione
di Ferrara, Ferrara, Italy}
\affiliation{Department of Physics, University of California,
Santa Cruz, California 95064}

\author{A.~P.~Young}
\affiliation{Department of Physics, University of California,
Santa Cruz, California 95064}

\date{\today}

\begin{abstract}
We determine the complexity of several constraint satisfaction problems using
the heuristic algorithm, WalkSAT. At large sizes $N$, the complexity increases
exponentially with $N$ in all cases. Perhaps surprisingly, out of all the
models studied, the hardest for WalkSAT is the one for which there is a
polynomial time algorithm.
\end{abstract}
\maketitle

\section{Introduction}
\label{sec:intro}

There is considerable interest, in different fields of science, in finding
efficient methods to solve
optimization problems. For a few such problems there are clever algorithms
which enable one to find the solution, for all cases, in a time which only
grows with a power of the size $N$ of the problem. Problems with such a
polynomial time algorithm are said to be in the complexity class P. In many
cases of interest, however, no polynomial time algorithm is known, though a
solution, if given, can be \textit{verified} in polynomial time. 
Decision problems for which
which ``yes" instances can be verified in polynomial time are
said to be~\cite{Garey:97}
in complexity class NP. There is no proof that P $\ne$ NP, though it is
generally assumed that they are different, i.e.~there exist problems which
cannot be solved in polynomial time algorithm, at least
in the worst case. There is a subset of
NP problems which have the property that any problem in NP can be mapped
into them in polynomial time. These are called NP-complete~\cite{Garey:97}.
Consequently if a polynomial time algorithm could be found for \textit{one}
NP-complete problem, all problems in NP could be solved in polynomial time and
so we would have P = NP (which, as stated above, is felt to be unlikely).

Several heuristic algorithms to solve optimization problems have been
proposed. (``Heuristic" means that the answer provided is not guaranteed to
be exact. This is in contrast to ``complete'' algorithms which guarantee to
find the solution, if one exists.) One well known example is simulated
annealing~\cite{kirkpatrick:83} (SA) in which an artificial temperature is
introduced and gradually set to zero. Another popular algorithm is
WalkSAT~\cite{WalkSAT} which is similar in spirit to simulated annealing
in that both make moves which reduce the ``energy'', but also sometimes make
moves which increase it to avoid being trapped in the nearest local minimum.

It has also been proposed to solve optimization problems on a quantum computer
using the quantum adiabatic algorithm (QAA)~\cite{farhi:01}, which is based on
quantum annealing~\cite{kadawoki:98}. To assess whether a quantum computer
could solve optimization problems more efficiently than a classical computer,
it is valuable to compare the efficiency of the QAA to solve a range of
optimization problems with that of classical heuristic algorithms, in
particular to see whether those problems which are harder classically are also
harder quantum mechanically.

As part of this project we report here results of the efficiency of a
classical algorithm to study several optimization problems of the ``constraint
satisfaction" type.  We chose WalkSAT rather than SA because the
implementation is simpler since there is just one parameter (the strength of
the ``noise") whereas in SA one has to decide on the whole annealing schedule
of temperature against time.  It is not obvious how to choose the best
annealing schedule, and different choices could lead to significant
differences in the number of sweeps needed to find the ground state, which
would be a disadvantage for us.  Also, there is a publicly available code for
WalkSAT which can be downloaded and easily compiled on different machines.

We study three NP-complete models and  two
versions of one model that is in the P class (known
as XORSAT). For
WalkSAT, all models we study require a computer time which increases exponentially with
$N$. Some are harder than others, in that the coefficient of $N$ in the
exponent is larger. Curiously the hardest of all is XORSAT even though there
exists a polynomial time algorithm for this problem. Hence, the difficulty of a
problem using a heuristic algorithm does not, in general, depend on whether it
is in the P or NP complexity class.
Interestingly, it is also known that XORSAT is very hard using the
QAA~\cite{jorg:09}.

The plan of this paper is as follows. Section~\ref{sec:models} describes the
four models that will be studied. Results are presented in Sec.~\ref{sec:results}
and our conclusions summarized in Sec.~\ref{sec:conclusions}.

\section{Models}
\label{sec:models}

We shall consider problems of the ``constraint satisfaction'' type, in which
there are $N$ bits
(or equivalently Ising spins)
and $M$ ``clauses'' where
each clause is
a logical condition on a small number of randomly chosen bits. A
configuration of the bits is a
``satisfying assignment'' if it satisfies all the clauses. Frequently,
in statistical mechanics approaches to these problems, one converts each
clause to an energy function, which depends on the bits in the clause, such
that the energy is zero if the clause is satisfied and is positive if it is
not. However, we will not need to do this here since WalkSAT uses the logical
structure of the clauses rather than an energy function.

Clearly, it is easy to satisfy all clauses if the ratio $\alpha \equiv M/N$
is small enough. In fact one expects an exponentially large number of
satisfying assignments in this region.
Conversely, if $M/N$ is very large, with high probability there will be
a conflict between different clauses. 
Hence there is a ``satisfiability transition'' at some value $\alpha_s$ where the
number of satisfying assignments goes to zero.  It is believed that it is
particularly hard to solve satisfiability problems close to the
transition~\cite{kirkpatrick:94}, and so we will work in this region.
Furthermore, when studying the efficiency of the QAA
numerically~\cite{farhi:01,young:08,young:10}, it is convenient
to consider instances with a
unique satisfying assignment (USA). Since we intend eventually to compare the
results presented here with results for the QAA, here we will also only
consider instances with a USA (which of courses, forces the system to be
close the transition).

We now discuss the different models that will be investigated in this paper.

\subsection{Exact Cover (Unlocked 1-in-3 SAT)}
\label{sec:unlocked}
Several numerical studies of the QAA~\cite{farhi:01,young:08,young:10}
consider the ``Exact Cover'' problem, in
which each clause consists of three bits chosen randomly, and the clause is
satisfied if one bit is one and the others are zero. In order to increase the
probability of a USA, Young et al.~\cite{young:08,young:10} removed isolated
bits, and clauses which are only connected to the others by one bit. Here we
use exactly the same instances as in Refs.~\cite{young:08,young:10}. For each
value of $N$, the number of clauses $M$ is chosen to maximize the probability
of a USA. The
values of $N$ and $M$ are shown in Table~\ref{tab:EC}.
It is found that the probability of a USA decreases with $N$, apparently
exponentially.

\begin{table}
\begin{center}
\begin{tabular}{|r|r|r|r|r|r|r|r|}
\hline\hline
N        & 16     & 32     & 64     & 128    & 192    & 256 \\
\hline
M        & 12     & 23     & 44     & 86     & 126    & 166 \\
\hline
$\alpha$ & 0.7500 & 0.7188 & 0.6875 & 0.6719 & 0.6563 & 0.6484 \\
\hline\hline
\end{tabular}
\end{center}
\caption{The number of clauses $M$ and number of bits $N$ used in the study of
the Exact Cover problem, from Ref.~\cite{young:08}.
The ratio $M/N$ is
expected to approach the value at the
satisfiability phase transition $\alpha_s \simeq 0.625$~\cite{knysh:04,raymond:07} for $N
\to \infty$. }
\vspace{-0.5cm}
\label{tab:EC}
\end{table}

This Exact Cover problem is sometimes called 1-in-3 SAT, for obvious reasons.
In subsections~\ref{sec:locked_1-in-3} and \ref{sec:locked_2-in-4},
we will discuss models with a special property called
``locked'', defined by Zdeborov\'a and
M\'ezard~\cite{zdeborova:08a,zdeborova:08b}. To distinguish the
present model from the locked models, we will refer to it as ``unlocked 1-in-3
SAT" from now on.

\subsection{Locked 1-in-3 SAT}
\label{sec:locked_1-in-3}

Recently Zdeborov\'a and
M\'ezard~\cite{zdeborova:08a,zdeborova:08b} have proposed that it is useful to
study a set of models, which they call ``locked'', which have the following
two properties:
\begin{enumerate}
\item
Every variable is in at least two clauses.
\item
Whether or not a clause is satisfied only depends on the sum of the bits in
it (occupation problem), and two successive values of the sum are not allowed.
Thus 1-or-3-in-4 is allowed, but 1-or-2-in-4 is not.
\end{enumerate}

It follows that one can not get from one satisfying assignment to another by
flipping a single bit. In fact, Zdeborov\'a and
M\'ezard argue that typically order $\ln N$ bits needs to be flipped.
They also argue that, locked instances are analytically ``simple'' (or at
least simpler than previously studied models such as random K-SAT) but are
computationally hard. They are therefore eminently suitable as benchmarks.

If the sites are chosen at random to form the clauses, the distribution of the
degree of the sites (i.e.~the number of clauses involving a site) would be
Poissonian. However, locked instances have a minimum degree of two, so instead we
use a truncated Poissonian distribution~\cite{zdeborova:08b} which is
Poissonian except that the probabilities for zero and one are set to zero.
We fix the ratio $M/N$ to be the critical value for the satisfiability
transition. According to Table I of Ref.~\cite{zdeborova:08b}, this is equal to
$\alpha_s = 0.789$. Since $M$ has to be an integer we take $M$ to be the
nearest integer to $\alpha_s N$.

Having generated these instances we run them through a (complete) 
Davis-Putnam-Logemann-Loveland (DPLL) ~\cite{davis:60,davis:62} code to
select those with a unique satisfying assignment (USA). 

\begin{figure}
\begin{center}
\includegraphics[width=\figurewidth]{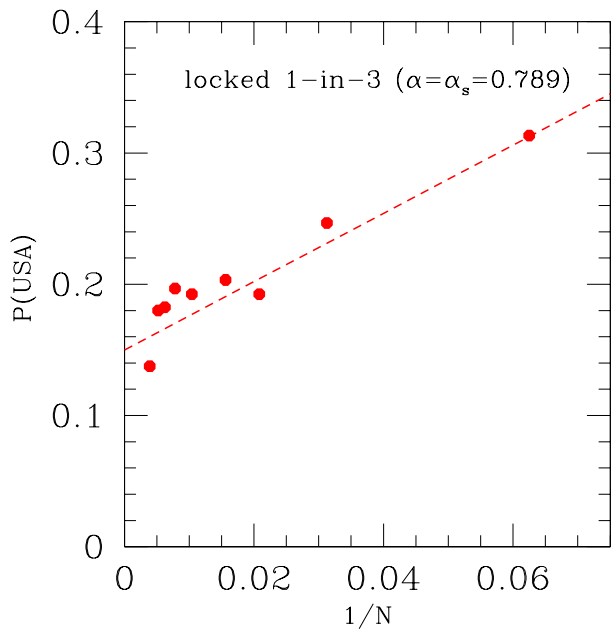}
\caption{(Color online) 
Probability of a unique satisfying assignment (USA) as a function of $1/N$ for
locked instances of 1-in-3 SAT at the satisfiability threshold. The line is a
guide to the eye.
}
\label{P_USA_1-in-3}
\vspace{-0.7cm}
\end{center}
\end{figure}

The probability of a USA only decreases slowly with $N$ and may tend to a
non-zero value as $N \to \infty$, see
Fig.~\ref{P_USA_1-in-3}. This
is in contrast to the unlocked instances in Sec.~\ref{sec:unlocked} for which
the probability decreases exponentially with $N$. The locked problem
therefore has the advantage that 
instances with a USA
should be a good representation of \textit{randomly chosen}
instances. 

\subsection{Locked 2-in-4 SAT}
\label{sec:locked_2-in-4}

We also consider locked 2-in-4 instances, in which a clause has four bits,
and a clause is satisfied if two are zero and two are one. Unlike the other models
discussed in this paper, this one has a symmetry under flipping all the bits.

\begin{figure}
\begin{center}
\includegraphics[width=\figurewidth]{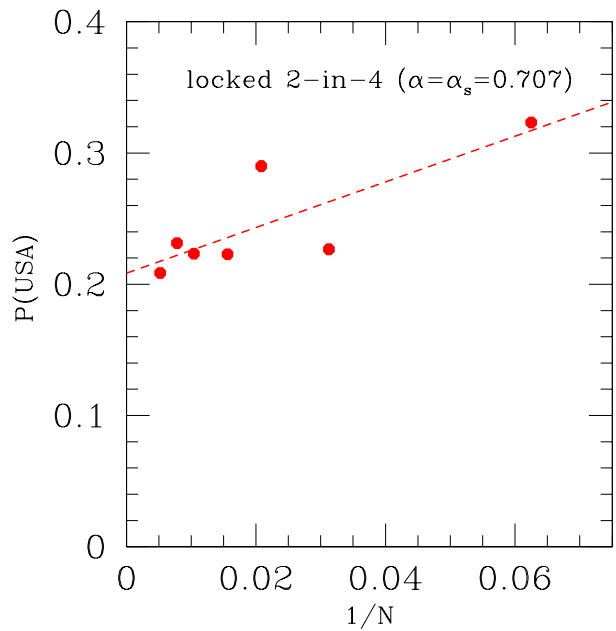}
\caption{(Color online) 
Probability of a unique satisfying assignment (USA) as a function of $1/N$ for
locked instances of 2-in-4 SAT at the satisfiability threshold. The line is a
guide to the eye.
}
\label{P_USA_2-in-4}
\vspace{-0.7cm}
\end{center}
\end{figure}

We fix the ratio $M/N$ to be the critical value for the satisfiability
transition. According to Table I of Ref.~\cite{zdeborova:08b}, this is equal to
$\alpha_s = 0.707$. Since $M$ has to be an integer we take $M$ to be the
nearest integer to $\alpha_s N$. As with the locked 1-in-3 instances in
Sec.~\ref{sec:locked_1-in-3} the probability of a USA seems to tend to a
non-zero value for $N \to \infty$, see Fig.~\ref{P_USA_2-in-4}.

\subsection{XORSAT}
\label{sec:XORSAT}

The exclusive-or of a set of bits is their sum (mod 2).  In the K-XORSAT problem,
K bits are chosen to form a clause and the clause is satisfied if their sum
(mod 2) is a specified value (either 0 or 1). Again, the problem to be
solved is whether there is an assignment of the
$N$ bits which satisfies all $M$ clauses. In fact, since the problem just
involves linear algebra (mod 2) the satisfiability problem can be solved in
polynomial time using, for example, Gaussian elimination. However, if there is
\textit{not} a satisfying assignment, there is no known polynomial time
algorithm to determine the minimal number of
unsatisfied clauses,
a problem known as MAX-XORSAT.

For XORSAT
instances with a USA, it is not difficult to show that one can gauge
transform any instance into one in which the sum of the bits of every clause
is equal to 0 (mod 2).
The USA is then all bits equal to 0,
(a ``ferromagnetic" ground state in statistical physics language).
Although this 
ground state is ``trivial", we shall see that it is very hard to find using a heuristic
algorithm.

Here we will take the case of $K = 3$, and study two variants of the model.

\subsubsection{3-regular 3-XORSAT}
Firstly, we will follow J\"org et
al.~\cite{jorg:09} in taking the ``3-regular'' case where every bit is in
exactly three clauses, a model which turns out to
be \textit{precisely} at the satisfiability
threshold. We denote this model as 3-regular 3-XORSAT.
As usual, we consider instances with a USA. Fortunately,
these are \textit{a non-zero
fraction}, about 0.285~\cite{jorg:09}, of the total, so the USA
instances should be a good representation of randomly chosen ones.
Note that $M = N$ for this
model.

\subsubsection{3-XORSAT}
Secondly, following the suggestion of Lenka
Zdeborov\'a, we will consider instances in which the distribution
of the number of clauses attached to a bit is truncated Poissonian with
\textit{mean} degree three.
For this model, which we denote as 3-XORSAT, we find numerically that
the fraction of instances with a
USA is also non-zero, about 0.25.
As usual, we restrict our attention to these instances. 

\section{Results}
\label{sec:results}

We study the models in Sec.~\ref{sec:models} using the WalkSAT~\cite{WalkSAT}
algorithm.  WalkSAT picks at random a clause which is currently unsatisfied
and flips a variable in that clause. With some probability the variable is
chosen to be the one which causes the fewest previously satisfied clauses to
become unsatisfied, and otherwise it is chosen at random. The probability that
the variable is chosen at random is called the ``noise parameter".  We know
that there is a (unique) satisfying assignment for each instance (since we
selected instances to have this condition using a DPLL algorithm).
We determine how many elementary WalkSAT ``flips'' are needed to find it, for
the different models for a range of sizes.

The overall logical condition which must be satisfied is the 
logical AND of each
clause. Most algorithms for solving satisfiability problems, including
WalkSAT, require that the
problem is expressed in
conjunctive normal form (cnf), in which each clause is written entirely 
in terms of
logical OR's. Note that OR just excludes one possible state of the bits.
The cnf representation of the problems studied
here is not unique. For example, in 1-in-3 SAT, a clause requires $x_1 + x_2 + x_3 =
1$, i.e.~one of the bits is 1 and the others are 0 (so there are 3 allowed
configurations).
This has to be written as a
logical AND of \textit{several} cnf clauses (each of which is comprised of
OR's). A natural choice is to
use five clauses as
\begin{align}
&(x_1 \lor x_2 \lor x_3) \land (\lnot x_1 \lor \lnot x_2 \lor \lnot x_3) \land \\
& (x_1 \lor \lnot x_2 \lor \lnot x_3) \land (\lnot x_1 \lor x_2 \lor \lnot x_3) \land \\
& (\lnot x_1 \lor \lnot x_2 \lor x_3)  \, ,
\end{align}
in which each clause disallows one of the $8-3\ (= 5)$ forbidden configurations.
However, we can actually combine two of the clauses together using a clause
with only two variables, for example as
\begin{align}
&(x_1 \lor x_2 \lor x_3) \land (\lnot x_1 \lor \lnot x_2 ) \land \\
& (x_1 \lor \lnot x_2 \lor \lnot x_3) \land (\lnot x_1 \lor x_2 \lor \lnot x_3)
\, .
\end{align}
We always chose the cnf representation that used the smallest number of
clauses.

\begin{figure}
\begin{center}
\includegraphics[width=\figurewidth]{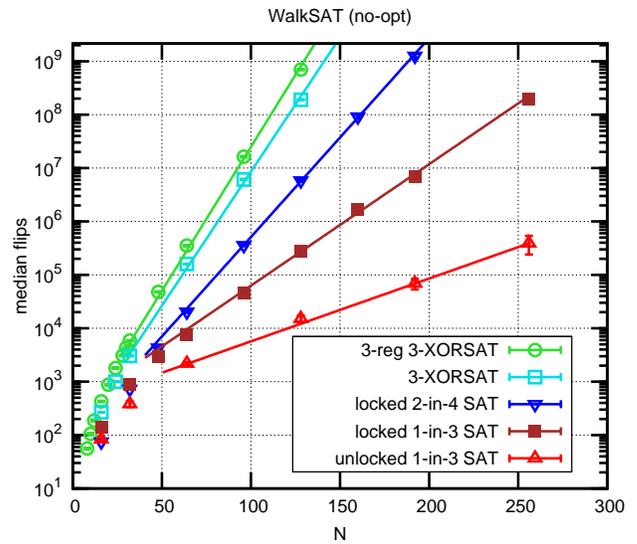}
\caption{(Color online) 
A log-lin plot of the median number of 
flips needed to solve the various problems as
a function of $N$ using WalkSAT with the default noise parameter. The straight
lines are fits to Eq.~\eqref{Nflip} for the larger sizes.
}
\label{all_no_opt_log-lin}
\end{center}
\end{figure}

First, we show results using the default value of the noise parameter (0.5).
Figure~\ref{all_no_opt_log-lin} plots the median
number of ``flips'' to find a solution as a function of $N$ for the different
models.
Note the logarithmic vertical scale. The data fits a straight line at large $N$
indicating a number of flips (which is proportional to the CPU time)
increasing exponentially, as expected.

\begin{table}
\begin{center}
\begin{tabular}{|c|c|}
\hline\hline
model        & $\mu$ \\
\hline
unlocked 1-in-3 SAT      & 0.027 \\
locked 1-in-3 SAT        & 0.052 \\
locked 2-in-4 SAT        & 0.086 \\
3-XORSAT                 & 0.116 \\
3-regular 3-XORSAT       & 0.124 \\
\hline\hline
\end{tabular}
\end{center}
\caption{
The rate of exponential increase of the number of flips, according to
Eq.~\eqref{Nflip} for the different
models using the default value of the noise.
}
\vspace{-0.5cm}
\label{tab:mu}
\end{table}

\begin{figure}
\begin{center}
\includegraphics[width=\figurewidth]{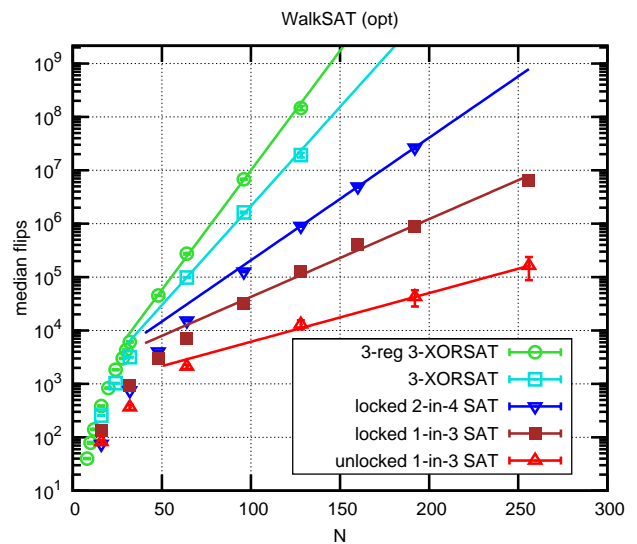}
\caption{(Color online) 
A log-lin plot of the median number of flips needed to solve the various problems as
a function of $N$ using WalkSAT with optimized noise parameters. The straight
lines are fits to Eq.~\eqref{Nflip} for the larger sizes. The results are
quite similar to those for the default noise shown in Fig.~\ref{all_no_opt_log-lin}
}
\label{all_opt_log-lin}
\end{center}
\end{figure}

We see that the easiest model is unlocked 1-in-3 SAT (Exact Cover), while the
hardest is 3-XORSAT. Both versions of 3-XORSAT are harder than any of the
other problems that we looked at, with the 3-regular version being somewhat
harder than the version with a Poissonian degree distribution. 

Writing the median number of flips as
\begin{equation}
\label{Nflip}
N_\text{flips} = A e^{\mu N}
\end{equation}
we present the values of $\mu$ in Table~\ref{tab:mu}.
It is curious that the one problem in the set (XORSAT) which is
in the polynomial time complexity class, P,  is the hardest for a heuristic
algorithm.

We have also investigated the extent to which WalkSAT can be improved by
optimizing with respect to the noise parameter. For each size, and each
instance, we try different values of the noise and home in on the value which
minimizes the number of
flips. Results for the median number of flips are shown in
Fig.~\ref{all_opt_log-lin}. Naturally, the numbers are a bit smaller than for
the unoptimized case in Fig.~\ref{all_no_opt_log-lin}, but we still see an
exponential growth at large sizes, and the ordering of the difficulty of the
different models is the same (e.g.~unlocked 1-in-3 SAT is the easiest and and
the two versions of 3-XORSAT are the hardest). The fits in Fig.~\ref{all_opt_log-lin}
correspond to the coefficients of $N$ in the exponential growth shown in 
Table~\ref{tab:mu_opt}, see Eq.~\eqref{Nflip}. 
Unsurprisingly, these are somewhat less than the
values for the unoptimized case shown in Table~\ref{tab:mu}. 

\begin{table}
\begin{center}
\begin{tabular}{|c|c|}
\hline\hline
model        & $\mu$ \\
\hline
unlocked 1-in-3 SAT      & 0.021 \\
locked 1-in-3 SAT        & 0.034 \\
locked 2-in-4 SAT        & 0.053 \\
3-XORSAT                 & 0.085 \\
3-regular 3-XORSAT       & 0.107 \\
\hline\hline
\end{tabular}
\end{center}
\caption{
The rate of exponential increase of the number of flips, according to
Eq.~\eqref{Nflip} for the different
models using optimized values of the noise. These values for $\mu$
are somewhat
smaller than those for unoptimized noise, in Table~\ref{tab:mu}, but the
overall trend between the different models is the same.
}
\vspace{-0.5cm}
\label{tab:mu_opt}
\end{table}

\section{Conclusions}
\label{sec:conclusions}
We have studied the complexity of the WalkSAT algorithm for four
constraint satisfaction problems, three of them in the NP complexity class
and one in the P complexity class. All show exponential complexity for large
sizes. Curiously, the hardest problem for WalkSAT is the one in P.
It will be interesting, in future work, to compare the relative hardness
of these problems for the classical WalkSAT
algorithm that we found here, with their relative hardness when using 
the quantum adiabatic algorithm.


\begin{acknowledgments}
One of us (APY) thanks
Florent Krzakala, Lenka Zdeborov\'a, Francesco Zamponi, Eddie
Farhi and David Gosset
for helpful discussions. We also thank Zdeborov\'a and Farhi for their
comments on an earlier version of this paper.
This work is supported in part by the National
Security Agency (NSA) under Army Research Office (ARO) contract number
W911NF-09-1-0391, and in part by the National Science Foundation under Grant
No.~DMR-0906366. MG has been partially supported by the Doctoral Program of the
IUSS (University of Ferrara) and by INFN.

\end{acknowledgments}

\bibliography{refs}

\end{document}